# *In situ* and *non situ* surface x-ray scattering studies of in plane Cs$^+$ ordering in Helmholtz planes of Pt(111) surface

Yihua Liu$^{a,†}$, Anthony Reiter$^b$, Christian Cammarota$^b$, Tomoya Kawaguchi$^a$, Michael S. Pierce$^{b,*}$, Vladimir Komanicky$^c$, Hoydoo You$^{a,*}$

$^a$ *Materials Science Division, Argonne National Laboratory, Argonne, Illinois, 60439*
$^b$ *Rochester Institute of Technology, Department of Physics Rochester NY 14623*
$^c$ *Safarik University, Faculty of Sciences, Kosice 04154, Slovakia*

\* [hyou@anl.gov](mailto:hyou@anl.gov); [mspsps@rit.edu](mailto:mspsps@rit.edu)
† Current Address: Lam Research, Tualatin, OR, USA

**Abstract**

In-plane ordering of Cs$^+$ layers in Helmhotlz planes was studied on Pt(111) surface in 0.1 M CsF electrolyte solutions with synchtrotron surface x-ray scattering techniques. The ordering was measured in a new transmission cell, designed for *in situ* and *non situ* measurements and high-temperature sample annealing all in the cell without sample transfer steps. At −850 mV vs. Ag/AgCl, (2×2) in-plane scattering peaks were weak under *in situ* condition and grew rapidly under *non situ* condition as the surface emersed from the electrolyte. The models for the (2×2) structures are presented and differences between *in situ* and *non situ* conditions are discussed.

## 1. Introduction

The structure and motion of ions in electrochemical double layers (EDL) is key information in basic electrochemistry and electrocatalysis(Markovic 2013). EDL is traditionally described as diffuse distributions of cations and anions, proposed by Gouy and Chapman(Bard and Faulkner 2001) and further developed extensively over decades(Gurney 1935, Grahame 1947, Wagner and Ross 1983, Grassi, Daghetti et al. 1987, Trasatti 1987, Ardizzone, Fregonara et al. 1990, Daghetti, Romeo et al. 1993, Trasatti and Doubova 1995). The early development of the diffuse EDL models can be found in an extensive review(Parsons 1990). In recent years, synchrotron x-ray techniques have been extensively used for studies of chemical and electrochemical double layers: on the membrane-aqueous interfaces using x-ray standing wave technique (Bedzyk, Bommarito et al. 1990), on the solid/liquid interfaces using crystal truncation rod measurements (Lucas, Thompson et al. 2011), and on liquid/liquid interfaces using x-ray reflectivity techniques.(Luo, Malkova et al. 2006).

The EDL in previous studies was considered for the distribution of ions only along the direction of surface normal. The assumption is that the ions have no in-plane ordering and distributed randomly parallel to the surface. However, for potentials significantly away from the potential of zero charge (PZC), the charged ions, sometimes hydrated, can be strongly pulled toward the solid surface and form relatively dense layers of the ions. The dense layer of the ions with repulsive interactions due to the same charges may induce significant in-plane structures that have not been observed heretofore. This in-plane structure, the focus of this study, differs from those occurring in chemisorption accompanying faradaic charge transfer reactions. In the case of $Cs^+$ studied here, the cation does not specifically chemisorb even at the largest negative potential that we measured *in situ*. Yet, in-plane ordering peaks are identified under *in situ* as well as *non*

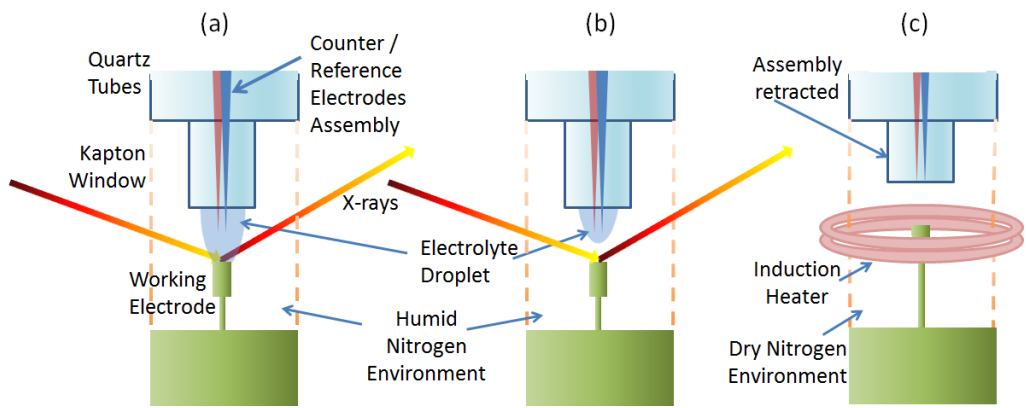

Figure 1. Schematic drawings of the transmission cell design: (a) The working electrode surface is immersed for electrochemical control and *in situ* transmission x-ray measurements. (b) The electrolyte droplet is lifted for *non situ* measurements. (c) The counter/reference electrode assembly is retracted and the electrode assembly was raised for the inductive annealing of the electrode.

*situ* (Stuve, Krasnopoler et al. 1995) conditions using the emersion technique(Kolb, Rath et al. 1983, Kolb 1987, Zurawski, Rice et al. 1987).

## 2. Experimental

### 2.1 Cell Design

A new transmission cell was designed and built for this experiment. In this cell, the sample surface can be annealed by an induction heater and immersed to an electrolyte without exposing the surface to ambient conditions. In this way, the pristine surface prepared is immediately used for electrochemical/x-ray measurements. The cell geometries are schematically shown in Figure 1. In (a), x-rays diffract from the surface through the electrolyte typically 2~3 mm thick. The x-ray transmission tends to produce large background scattering. The background scattering can be subtracted but the signal to noise ratio limits the detection of weak peaks. In (b), x-rays do not go through the water, therefore, the background scattering is much lower. The electrochemical control is lost and the situation is equivalent to the *non situ* condition in the UHV transfer experiments(Stuve, Krasnopoler et al. 1995). In (c), the inner quartz tube, filled with electrolyte and assembled with reference and counter electrodes, is retracted. Then, the Pt(111) crystal with the long Pt wire tail was raised to the height of the coil of the pre-aligned induction heater. This configuration is used for sample annealing.

*2.2 Synchrotron X-ray Measurements*

Synchrotron x-ray measurements were performed at 11ID-D beamline equipped with a '4S+2D' geometry six-circle diffractometer(You 1999) at Advanced Photon Source (APS). Pt(111) surface, precut and polished, had a miscut of <0.1°. The hexagonal (hex) index ($a^* = 4\pi\sqrt{2} / \sqrt{3}a$ and $c^* = 2\pi / \sqrt{3}a$ where a=3.9242Å) of face-centered cubic (fcc) structure was used in the experiments(Huang, Gibbs et al. 1990) where $(111)_{fcc}$, $(1\bar{1}1)_{fcc}$, and $(200)_{fcc}$ are indexed to $(003)_{hex}$, $(101)_{hex}$, and $(012)_{hex}$, respectively. 0.1 M CsF and CsCl electrolyte was prepared from the CsF and CsCl solid salt of 99.99% in metals-basis purity from Puratronic® dissolved in 18 MΩ·cm water. The results in CsCl are not reported here because the results were essential identical to those in CsF. The potential range used in the study was −850 mV to 400 mV vs. Ag/AgCl reference electrode. The counter electrode was a platinum wire. −850 mV was used for *in situ* $Cs^+$ structure measurements and also as the emersion potential for *non situ* measurements. The platinum crystal was annealed to ~1500 K before each set of experiments in dry Ar-3%$H_2$ inert gas flow. The Pt(111) crystal is cooled to room temperature and the gas flow is switched to bubbled humid $N_2$. The surface was checked for readiness and pre-oriented by x-ray reflectivity and Bragg diffraction before made in contact with the electrolyte droplet. In this way, the exposure of x-rays to the electrolyte was minimized. During the measurements, the x-ray shutter was open only for signal counting to minimize the x-ray exposure to the electrolyte. The open circuit potential can drift during x-ray exposure(Nagy and You 1995). Therefore, precautions were exercised and experiments were repeated with different intensities of incoming x-rays to ensure that the results presented here is in any way affected by the x-ray exposure.

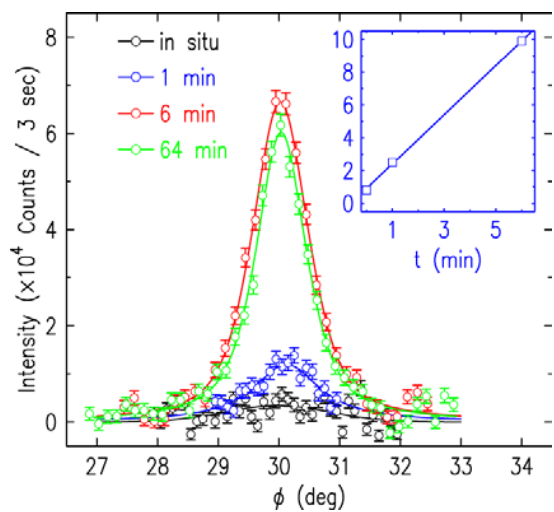

Figure 2. Scans at (0.5 0.5 0.33) *in situ* and 1, 6, 64 min *non situ* after the emersion. The inset shows the integrated intensities mesasured at 0 (in situ), 1, and 6 min.

## 3. Results and Discussion

*In situ* experiment was performed first. While the potential was held at −850 mV, various in-plane vector positions were scanned until a weak surface peak at (0.5 0.5 0.33) was identified. Then, the electrolyte droplet was withdrawn from the surface (Figure 1b) at −850 mV while holding the surface vertical for a quick emersion. The scan through (0.5 0.5 0.33) was immediately repeated and several successive scans are shown in Figure 2. The integrated intensities measured *in situ*, at 1 min, at 6 min, and at 64 min after the emersion were 0.9(1), 2.2(1), 8.9(1), and 8.5(1), respectively. The scan at open circuit was flat within the noise, not shown here for clarity. The intensity grows rapidly for the first 6 min after the emersion. It is important to note that the peak intensity goes back to the *in situ* condition if the surface is re-immersed immediately after the 1 min measurements. This indicates that the $Cs^+$ remains hydrated up to this point and the intensity increases mainly due to the lower background. If the surface is re-immersed after several min, however, the intensity remains strong and unresponsive to the applied potential, indicating that $Cs^+$ is dehydrated, at least partially, and possibly chemisorbed (or strongly adsorbed) to Pt(111) surface. In this case, the surface has to be reannealed to recover a clean surface. The intensity decreases eventually even under the humid $N_2$ flow in an hour after the emersion. The surface is no longer clean, probably due to the oxygen impurities in the cell. At this point, again, the surface has to be re-prepared to recover the pristine surface condition.

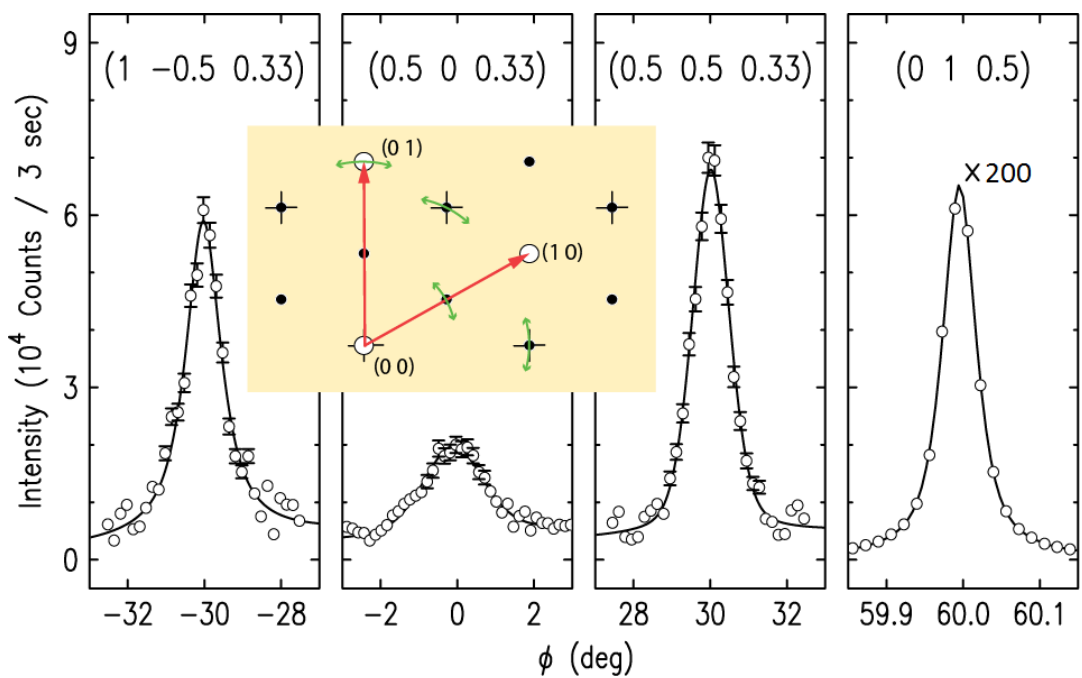

Figure 3. Three $Cs^+$ peaks and a Pt surface peak. (1 −0.5 0.33), (0.5 0 0.33), and (0.5 0.5 0.33) are the superlattice peaks due to the (2×2) $Cs^+$ layer and (0 1 0.5) is a Pt surface anti-Bragg peak. The inset shows a 2d in-plane reciprocal space. The green curved arrows indicate the directions of the scans.

The superlattice peak at (0.5 0.5 0.33) indicates a (2×2) structure. In order to determine the structure, three (2×2) peaks, (1 −0.5 0.33), (0.5 0 0.33), and (0.5 0.5 0.33), were measured. Pt(0 1 0.5) was also measured as a calibration point. These peaks are compared in Figure 3. The inset shows the reciprocal space map with the curved green arrows indicating the directions of the $\phi$ scans. Comparing the widths, the superlattice peaks are ~20 times broader than the Pt peak. The longitudinal scans (not shown) are also more than ~10 times broader. These scans indicate that the average domain size of the superlattice is in the order of a few tens nm. The intensities of these peaks all show little dependence on $L$ values, indicating that they are indeed from a monolayer structure. Pt(0 1 0.5), is the anti-Bragg peak between two Bragg peaks, (0 1 2) and (0 1 −1). The intensity of this peak can be estimated from an expression, $\left|\frac{f_{Pt}}{1-e^{i2\pi(L-1)}}\right|^2$, for (0 1 $L$) crystal truncation rod(Robinson 1986), where $f_{Pt}$ is the form factor for a platinum atom. The form factors for the diffraction angles concerned here are essentially the atomic numbers ($f_{Pt}$ = 78). The calculated intensity of (0 1 0.5) is then $[4(78/2)]^2$ for a (2×2) unit cell where 4 comes from 4 atoms in a unit cell.

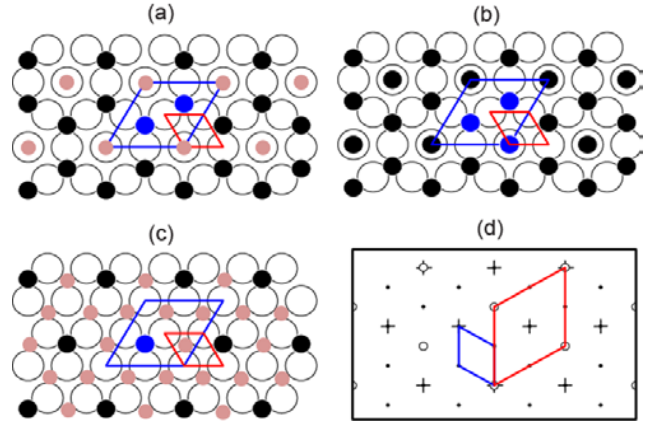

Figure 4. Three (2×2) models considered: (a) two sublattice unitcell, (b) three sublattice unitcell, (c) single sublattice unitcell. The open circles indicate Pt atoms, black solid circles represents Cs+ ions, and small pink circles show possible sites for water molecules. (d) shows the receiprocal space unitcells for Pt (1×1) (red) and Cs$^+$ (2×2) (blue). The circles are Pt receiprocal lattice and + and dots are the (2×2) superlattices.

In calculating the (2×2) superlattice peaks, three models shown in Figure 4 are considered. There are other possiblities. However, most of them can be considered as the variations or combinations of the three models by moving the Cs$^+$ positions to non-symmetric sites.

Therefore, only these three models will be considered: two sublattice model (a), three sublattice model (b), and a single sublattice model (c). Since $Cs^+$ ions are hydrated, water molecules are incoporated into the lattice. However, they are not included in the calculations because they are weak scatters. The calculated intensities for the three models and the experimental intensities are shown in Table 1. The all values are scaled by setting the intensity of (0 1 0.5) as unity.

The experimental intensities in Table 1 are from the *non situ* peaks shown in Figure 3. The intensities for (0.5 0.5 0.33) and (1 −0.5 0.33) are 10% and 9%, respectively, and that for (0.5 0 0.33) is 3%. The Debye-Waller (DW) factors of $Cs^+$, which are unknown, are not included. Therefore, the calculated intensities are the upper bounds and the the experimental intensities cannot be similar, if the $Cs^+$ layer is as well ordered as Pt layer, or larger than the calculated ones. This eliminates the single sublattice model (c). Among (a) and (b), (b) can easily eliminated because the measured intensity of (0.5 0 0.33) is small but not zero. In the case of model (a), the intensities make sense if the DW factor significantly reduced the intensity. The $Cs^+$ ions are expected to be quite disordered with significant domain boundaries because $Cs^+$ domain sizes are much smaller than the Pt(111) surface domain size as discussed in Figure 3. The DW factor is $e^{-(\sigma q)^2/3}$, where $q=4\pi\sin(\theta)/\lambda$ and $\sigma$ is the mean squareed displacement. The 80% reduction of the DW factor suggests $\sigma = $ ~0.9 Å, which is ~33 % of the Pt-Pt distance. The measured intensity ratio between (0.5 0.5 0.33) and (0.5 0 0.33) also agrees with the calaculated ratio.

Table 1. The comparison of the calculated (2×2) intensities to the measured *non situ* and *in situ* intensities for the models shown in Figure 4. The *in situ* intensities are measured in 1 min after emersion. All intensities are normalized by (0 1 0.5) intensity.

|                  | (a)  | (b)  | (c)  | *Non situ* intensity | *In situ* intensity |
|------------------|------|------|------|----------------------|---------------------|
| (0.5 0.5 0.33)   | 0.50 | 1.12 | 0.09 | 0.10(1)              | 0.03(1)             |
| (1. −0.5 0.33)   | 0.50 | 1.12 | 0.09 | 0.9(1)               | 0.03(1)             |
| (0.5 0.0 0.33)   | 0.12 | 0    | 0.09 | 0.03(1)              | 0.03(1)             |

For the *in situ* condition, the (0.5 0.5 0.33) intensity (Figure 2) is barely above the background. The *in situ* intensity of (0.5 0 0.33) is also close to the background (not shown). In the first scan immediately after emersion, the (0.5 0.5 0.33) intensity is ~3% of the (0 1 0.5) intensity or ~25% of the full *non situ* intensity. The delay time for the first scan is about a minute, which includes the time for withdrawing the electrolyte, interlocking the door of the x-ray hutch, and scanning the peak. Likewise, the (0.5 0 0.33) intensity, measured immediately following the first (0.5 0.5 0.33) measurement (in ~2 min after the emersion), is again ~3% of the (0 1 0.5) intensity. Note that the *non situ* (0.5 0 0.33) intensity does not change over time while the *non situ* (0.5 0.5 0.33) and (1 −0.5 0.33) peaks grow in time (Figure 2). The integrated intensities measured within 2 min indicate that the structures are different between the *non situ* and *in situ* conditions. It suggests that additional $Cs^+$ ions must be incorporated into the (2×2) structure during the first several minutes after emersion, probably from the thin electrolyte layer invisible yet still remaining after the emersion. Therefore, the *in situ* (2×2) structure should be close to the model (c) where the intensities of (0.5 0.5 0.33), (1 −0.5 0.33), and (0.5 0 0.33) are all weak and similar each other. Since the calculated intensities of the model (c) are 9% each for (0.5 0.5 0.33), the ~3% is reasonable for the model (c). This is also consistent with our recent crystal truncation rod measurements of the same system{Kawaguchi, 2017 #869}{Liu, 2017 #1029} where the direct inversion technique was used to obtain the $Cs^+$ peak density of ~0.7 $e^-/Å^3$ at 3.5 Å distance above the Pt(111) surface. Under this scenario, the highly disordered hydrated $Cs^+$ ions maintain the short-range single-sublattice (2×2) model (c) structure in electrolyte. As the solution thins in several min after emersion, however, additional Cs+ ions are incorporated into the lattice to form the two-sublattice model (a) structure.

## 4. Conclusions

In situ and non situ studies of $Cs^+$ cations in Helmholtz planes were performed on Pt(111) surface in 0.1 M CsF electrolyte using synchrotron in-plane x-ray diffraction. In both conditions, (2×2) peaks were observed. The in situ (0.5 0.5) peak intensity was weak and consistent with the simple (2×2) structure with a single sublattice occupied by $Cs^+$. The non situ (0.5 0.5) peak was strong, consistent with the model where two $Cs^+$ sublattices are occupied. These observations lead to two important conclusions. First, the hydrated cations can order in the first Helmholtz plane under the large polarization. This is true at least for hydrated $Cs^+$ layer in the Helmholtz planes. The hydrated ions are likely not registered to the Pt(111) substrate while the chemisorbed ions are registered. Second, the structures in non situ condition can be different from those in in situ condition. The non situ structures provide still useful information about the electrochemical surface. However, the hydration of the cations may change and the surface density of the cations can increase because of dehydration during and after the emersion process.

**Acknowledgements**

The work was supported by the U.S. Department of Energy (DOE), Office of Basic Energy Science (BES), Materials Sciences and Engineering Division and use of the APS by DOE BES Scientific User Facilities Division, under Contract No. DE-AC02-06CH11357. One of the authors (TK) thanks the Japanese Society for the Promotion of Science (JSPS) for JSPS Postdoctoral Fellowships for Research Abroad. The work at RIT was supported by the Research Corporation for Science Advancement (RCSA) through a Cottrell College Science.

# References

Ardizzone, S., G. Fregonara and S. Trasatti (1990). "INNER AND OUTER ACTIVE SURFACE OF RUO2 ELECTRODES." Electrochimica Acta **35**(1): 263-267.

Bard, A. J. and L. R. Faulkner (2001). Electrochemical methods : fundamentals and applications. New York, John Wiley.

Bedzyk, M. J., G. M. Bommarito, M. Caffrey and T. L. Penner (1990). "DIFFUSE-DOUBLE LAYER AT A MEMBRANE-AQUEOUS INTERFACE MEASURED WITH X-RAY STANDING WAVES." Science **248**(4951): 52-56.

Daghetti, A., S. Romeo, M. Usuelli and S. Trasatti (1993). "SINGLE-ION ACTIVITIES BASED ON THE ELECTRICAL DOUBLE-LAYER MODEL - AN INDIRECT TEST OF THE GOUY-CHAPMAN THEORY." Journal of the Chemical Society-Faraday Transactions **89**(2): 187-193.

Grahame, D. C. (1947). "THE ELECTRICAL DOUBLE LAYER AND THE THEORY OF ELECTROCAPILLARITY." Chemical Reviews **41**(3): 441-501.

Grassi, R., A. Daghetti and S. Trasatti (1987). "APPLICATION OF THE GOUY-CHAPMAN-STERN-GRAHAME MODEL OF THE ELECTRICAL DOUBLE-LAYER TO THE DETERMINATION OF SINGLE ION ACTIVITIES OF KF AQUEOUS-SOLUTIONS." Journal of Electroanalytical Chemistry **226**(1-2): 341-349.

Gurney, R. W. (1935). "Theory of electrical double layers m adsorbed films." Physical Review **47**(6): 479-482.

Huang, K. G., D. Gibbs, D. M. Zehner, A. R. Sandy and S. G. J. Mochrie (1990). "PHASE-BEHAVIOR OF THE AU(111) SURFACE - DISCOMMENSURATIONS AND KINKS." Physical Review Letters **65**(26): 3313-3316.

Kolb, D. M. (1987). "UHV TECHNIQUES IN THE STUDY OF ELECTRODE SURFACES." Zeitschrift Fur Physikalische Chemie Neue Folge **154**: 179-199.

Kolb, D. M., D. L. Rath, R. Wille and W. N. Hansen (1983). "AN ESCA STUDY ON THE ELECTROCHEMICAL DOUBLE-LAYER OF EMERSED ELECTRODES." <u>Berichte Der Bunsen-Gesellschaft-Physical Chemistry Chemical Physics</u> **87**(12): 1108-1113.

Lucas, C. A., P. Thompson, Y. Gruender and N. M. Markovic (2011). "The structure of the electrochemical double layer: Ag(111) in alkaline electrolyte." <u>Electrochemistry Communications</u> **13**(11): 1205-1208.

Luo, G. M., S. Malkova, J. Yoon, D. G. Schultz, B. H. Lin, M. Meron, I. Benjamin, P. Vanysek and M. L. Schlossman (2006). "Ion distributions near a liquid-liquid interface." <u>Science</u> **311**(5758): 216-218.

Markovic, N. M. (2013). "ELECTROCATALYSIS Interfacing electrochemistry." <u>Nature Materials</u> **12**(2): 101-102.

Nagy, Z. and H. You (1995). "RADIOLYTIC EFFECTS ON THE IN-SITU INVESTIGATION OF BURIED INTERFACES WITH SYNCHROTRON X-RAY TECHNIQUES." <u>Journal of Electroanalytical Chemistry</u> **381**(1-2): 275-279.

Parsons, R. (1990). "ELECTRICAL DOUBLE-LAYER - RECENT EXPERIMENTAL AND THEORETICAL DEVELOPMENTS." <u>Chemical Reviews</u> **90**(5): 813-826.

Robinson, I. K. (1986). "CRYSTAL TRUNCATION RODS AND SURFACE-ROUGHNESS." <u>Physical Review B</u> **33**(6): 3830-3836.

Stuve, E. M., A. Krasnopoler and D. E. Sauer (1995). "RELATING THE IN-SITU, EX-SITU, AND NON-SITU ENVIRONMENTS IN SURFACE ELECTROCHEMISTRY." <u>Surface Science</u> **335**(1-3): 177-185.

Trasatti, S. (1987). "APPLICATION OF THE GOUY-CHAPMAN-STERN-GRAHAME MODEL OF THE ELECTRICAL DOUBLE-LAYER TO 2 UNUSUAL CASES." <u>Journal of the Electrochemical Society</u> **134**(3): C137-C137.

Trasatti, S. and L. M. Doubova (1995). "CRYSTAL-FACE SPECIFICITY OF ELECTRICAL DOUBLE-LAYER PARAMETERS AT METAL/SOLUTION INTERFACES." <u>Journal of the Chemical Society-Faraday Transactions</u> **91**(19): 3311-3325.

Wagner, F. T. and P. N. Ross (1983). "THICKNESS OF ELECTROLYTE LAYERS ON EMERSED PT ELECTRODES." Journal of the Electrochemical Society **130**(8): 1789-1791.

You, H. (1999). "Angle calculations for a '4S+2D' six-circle diffractometer." Journal of Applied Crystallography **32**: 614-623.

Zurawski, D., L. Rice, M. Hourani and A. Wieckowski (1987). "THE INSITU PREPARATION OF WELL-DEFINED, SINGLE-CRYSTAL ELECTRODES." Journal of Electroanalytical Chemistry **230**(1-2): 221-231.